\documentclass[12pt]{iopart}
\newcommand{\be}{\begin{equation}}
\newcommand{\ee}{\end{equation}}
\begin{document}

\title[A solution of the van Dam-Veltman-Zakharov discontinuity problem]{A solution of the van Dam-Veltman-Zakharov discontinuity problem in the
frame of the Poincar\'{e}-covariant field gravitation theory}

\author{Yurij Baryshev$^1$  and  Sergey Oschepkov$^2$}

\address{$^1$Astronomical Department of the Saint Petersburg State University, Universitetskij pr.28, Staryj Peterhoff,
St. Petersburg, 198504 Russia

$^2$V.I. Vernadsky Crimean Federal University, prospekt Vernadskogo 4 , Simferopol, 295007 Russia
}

\ead{yubaryshev@mail.ru; y.baryshev@spbu.ru; sao@cfuv.ru}

\begin{abstract}
The  van Dam-Veltman-Zakharov (vDVZ) mass discontinuity problem can be solved
in the frame of the linear approximation of the Poincar\'{e}-covariant ($M_4, \eta^{\mu\nu}$) second rank symmetric tensor field ($\psi^{\mu\nu}$) gravitation theory.
Conservation of the source energy-momentum tensor  $T^{\mu\nu}{}_{,\,\nu}=0$,
together  with gauge invariance of the field equations, lead to generation of two
intrinsic irreducible non-ghost dynamical fields:
4-traceless symmetric tensor $\phi^{\,\mu\nu}$ (spin-2 universal attraction) and
4-trace $\psi = \eta_{\mu\nu}\psi^{\mu\nu}$ (spin-0 universal repulsion).
Due to difference in the signs of these forces the total free field Lagrangian
contains different signs for the tensor and scalar dynamical fields.
Generalized Fierz-Pauli mass term in total spin-2 $\bigoplus$ spin-0
Lagrangian gives natural massless limit for
$ m_g \rightarrow 0$, so the mass discontinuity paradox is absent.
The Newtonian gravity and relativistic gravity effects, including positive localizable energy density of both parts of the gravitational field, are derived. Experimental test of the reality of the dynamical spin-0 repulsive field can be
performed by LIGO-Virgo gravitational wave observations.
\end{abstract}



\section{Introduction}
Recent direct detections of gravitational waves (GW) by LIGO-Virgo observatories
(reviewed in  \cite{abbott18}) have opened new epoch in fundamental physics.
Especially the questions about quantum nature of gravitational interaction
and  existence of the new massive gravitational wave polarizations
have profound consequences for physics and astrophysics \cite{deRham17}.


In classic papers, van Dam-Veltman-Zakharov (\cite{70vDV}, \cite{70Zakh}) studied a massive
spin-2 field $\psi_{\mu\nu}$  in the flat Minkowski space, that couples to
matter as the $\psi_{\mu\nu} T^{\mu\nu}$, i.e. as the effect of graviton
exchange between  two material objects ( $T^{\mu\nu}$  is the energy-momentum
tensor of any matter). They showed that, at distances
much smaller than the Compton wavelength of the massive graviton,
one recovers Newton's law by an appropriate choice of the spin-2 coupling constant.
However, in this small-mass limit, the bending angle of light by a massive
body approaches $3/4$ of the Einstein result of general relativity theory (GRT).
This is the van Dam-Veltman-Zakharov (vDVZ)
discontinuity paradox, which means that one must take the exact zero mass for
the spin-2 graviton
in the Fierz-Pauli type \cite{39FierzPauli} quantum field theories having GRT as weak field approximation.

In modern literature there are many suggestions how to avoid vDVZ paradox by means of constructing
different non-linear theories of massive gravity: higher-dimensional theories
leading to the Dvali-Gabadadze-Porrati model, cascading gravity,
ghost-free massive gravity, Lorentz-violating massive gravity, non-local
massive gravity and more (\cite{12Hinter}, \cite{14Rham}, \cite{03Damour}, \cite{08RubTin}).

Here we present another way for resolution of the vDVZ discontinuity problem
in the frame of the linear weak field approximation of the Poincar\'{e}-covariant
symmetric second rank tensor quantum field theory  in Minkowski space.
Conservation of the source energy-momentum tensor  $T^{\mu\nu}{}_{,\,\nu}=0$,
together  with gauge invariance of the field equations, lead to generation of two
intrinsic irreducible non-ghost dynamical fields:
4-traceless symmetric tensor $\phi^{\mu\nu}$ spin-2 attraction and
4-trace $\psi = \eta_{\mu\nu}\psi^{\mu\nu}$ spin-0 repulsion, which describe the quantum nature
of the gravitational interaction.

Due to difference in the signs of these forces the total free field Lagrangian
contains different signs for the tensor and scalar dynamical fields.
Decomposition of a reducible tensor onto irreducible parts is applied to both gravitational
potentials $\psi_{\mu\nu}$ and
the source $T_{\mu\nu}$ of this field, so as a result we get two non-ghost quantum
dynamical fields describing observed gravitational phenomena.
Decomposition of the massless symmetric tensor field onto two dynamical fields (spin-2 attraction and spin-0 repulsion) was suggested in
\cite{80SokBar}, \cite{97BarSok}.
Here we show that in the frame of the massive
linearised weak field Poincar\'{e}-covariant Feynman's
field gravity approach there is no mass discontinuity paradox.


\section{Poincar\'{e}-covariant field approach to gravitation}

Physical arguments for the \emph{Necessity of Gravitational Quantization}\,
and understanding quantum nature of the gravitational interaction
were presented by Richard Feynman many years ago
in his comments at Chapel Hill Conference (1957) \cite{57Feynman}, his
repot at  Warsaw Conference (1963) \cite{63Feynman}
and in lectures which he taught as a course at Caltech
on gravitation during the 1962-63 academic year, first published in 1995 \cite{95Feynman}.

Feynman suggested to consider the gravitational interaction as a non-metric
quantum field in Minkowski space, in addition to other fundamental physical
quantum interactions - QED, QEWD, QCD, which are based on global Poincar\'{e}
group (inhomogeneous Lorentz group).
He emphasized that
\emph{"the geometrical interpretation is not really necessary or
essential to physics"} and
\emph{"the fact is that a spin-two field has this
geometrical interpretation; this is not something readily explainable
$-$ it is just marvelous"} (\cite{95Feynman}, p.113).

The question about identity of the Field Gravity Theory (FGT) with the geometrical
General Relativity Theory (GRT) was debated for a long time
(e.g. \cite{95Feynman},  \cite{73MTW}, \cite{61Thirring}  ,
\cite{00Str}, \cite{08Pad}, \cite{08Mag}, \cite{17Bar}).
Indeed, in the weak field approximation all actually performed relativistic
gravity experiments have the same predicted values in both FGT and GRT
(reviewed in \cite{73MTW}, \cite{17Bar}).
The initial basic principles of GRT are:
fundamental role of the metric $g_{\mu\nu}$ of the curved space ($R_{\mu\nu}$),
equivalence of inertial and accelerated by gravity reference frames,
energy-momentum pseudotensor ($t_{\mu\nu}$) of gravitation field, non-trivial
topology for black hole solutions.

While FGT has conceptually different basic principles:
the tensor field gravitational potential $\psi_{\mu\nu}$
in the Minkowski space (with metric $\eta_{\mu\nu}$) and the Poicar\'{e}
transformation group,
which guaranty existence of the conserved energy-momentum tensor $T^{\mu\nu}_{(g)}$
of the gravitational field, localizable and positive energy density of static and
variable gravitational field ($T^{00}_{(g)}(x)>0$), and initial quantum nature of the gravitational interaction (gravity force due to exchange by gravitons).

Hence one has to investigate Feynman's quantum field approach more carefully for understanding the physical nature of gravity phenomenon.

\subsection{Action and Lagrangians}

So, let us consider the relativistic (Poincar\'{e}-covariant) quantum field approach
to gravitation as a theory of the second rank symmetric tensor field
$\psi_{\mu\nu}$ in Minkowski space
(\footnote{
	We use main definitions and notations similar to
	Bogolubov \& Shirkov \cite{80BogShir} (Quantum Fields 1980)
	so the Minkowski metric $\eta_{\mu\nu}$ has signature $(+,-,-,-)$,
	4-dimensional tensor indices are denoted by Greek  letters
	$\alpha,\beta,\mu,\nu...$
	which take on the values 0, 1, 2, 3, and 3-dimensional quantities
	are denoted by
	Latin letters $i,k,l...$ and take on the values 1,2,3.
	Derivatives are denoted by
	$\partial\phi/\partial x^{\nu} =
    \partial^{\nu}\phi = \phi\,,\nu\, $.
	We use  explicit dimensional values of the
	fundamental constants $G, \hbar, c$ and also
$\hbar = c =1$  system.
}).
We start from the first step in the iteration procedure, i.e. linear weak field,
which is described by
the action  $\mathcal{A}$ and the Lagrangians
$\Lambda (\psi^{\mu\nu}(x),\, \psi^{\mu\nu}_{\,\,\,\,,\,\sigma}(x)\,,\, T^{\mu\nu})$
for the system of gravitationally interacting  particles/fields, which  include the three parts:
\be \label{act-fgt}
\mathcal{A} = \mathcal{A}_{(\mathrm{g})} + \mathcal{A}_{(\mathrm{int})} +
\mathcal{A}_{(\mathrm{m})} = \frac{1}{c}
\int \left(\Lambda_{(\mathrm{g})} + \Lambda_{(\mathrm{int})} +
\Lambda_{(\mathrm{m})} \right) d\Omega\,.
\ee
The notations (g), (int), (m)
refer to the actions for the gravity field, the interaction, and
the matter (particles/fields), $d\Omega =cdt\, dV =d^4x$ .
The physical dimension of the action is
\be \label{act-dim}
[Action]= [energy\,\, density]\times[volume]\times[time] = [\hbar],
\ee
meaning that the measurable energy density of the
gravitational field should
be defined in the theory at the basic conceptual level
together with volume and time.
We shall use also the dimensionless form of the action:
\be \label{act-dim-less}
\mathcal{S} = \frac{\mathcal{A}}{\hbar}=
\frac{1}{c\hbar}
\int \left(\Lambda \right) d\Omega\,
=\frac{1}{e^2_{gPl}}
\int \left(\Lambda \right) dx^4 ,
\ee
where the Planck gravitational charge $e_{gPl}$ is
\be \label{e-Pl}
e_{gPl} = \sqrt{G}\,\,M_{Pl} ,
\quad
\mbox{\rm and}
\quad
M_{Pl} = \sqrt{\frac{\hbar c}{G}} .
\ee

In the field approach the principle of
\emph{universality of gravitational interaction} (UGI)
states that the gravitational field $\:\psi^{ik}$
interacts with all kinds of matter via their energy-momentum tensor
$T^{ik}$, so the Lagrangian for the interaction has the universal form:
\be \label{ugi}
\Lambda_{(\mathrm{int})} =  -
\frac{1}{c^{2}} \psi_{ik} T^{ik}
\ee
The principle of universality (\ref{ugi}) was introduced in \cite{50Mosh} and
it replaces the equivalence principle used in the geometrical
approach. From the principle of UGI
and the stationary action principle one can derive
those consequences of the equivalence principle, which
do not create paradoxes. An important consequence of UGI (\ref{ugi}) is that the
free fall acceleration of a body does not depend on its total mass,
but does depend on the direction and value of test particle velocity
(\cite{61Thirring}, \cite{02Bar}).
This is also consistent with the factor 2 (as $1+v^2/c^2)$ in the bending angle
of light by the Sun.

Dynamics of the field generated by the source and the motion of
the matter in the field are described by the equations of motion
derived from the \emph{stationary action principle}
$ \delta \mathcal{S} =0 $.
So assuming that the \emph{total EMT of matter is fixed} and varying
$\delta\psi^{\mu\nu}$ in (\ref{act-dim-less}) we get the field equations
in the form:
\be \label{field-eq}
\frac{\delta \mathcal{S}}{\delta\psi^{\mu\nu}(x)} =
\frac{\partial \Lambda}{\partial \psi^{\mu\nu}(x)} -
\frac{\partial}{\partial x^{\sigma}}
\frac{\partial \Lambda}{\partial
	\psi^{\mu\nu}_{\,\,\,\,,\sigma}(x)} = 0
\ee

\subsection{The connection between the signs of the Lagrangian,
	the field energy and the force}

From Lagrangians in action (\ref{act-fgt}), (\ref{act-dim-less}) it can be seen that
a field $\varphi_{att}$ interacting with matter via (\ref{ugi}) can
describe the universal attraction force $\vec{F}_{att}$ between two masses.
However if we change the sign  of the free field Lagrangian $\Lambda_{(g)}$
then both  the  potential and the force will change their signs
($\varphi_{rep} = - \varphi_{att}, \,\,\, \vec{F}_{rep} = -\vec{F}_{att} $ ),
hence we get universal repulsion.
The sign of the coupling constant makes no difference since it appears
as a square in diagrams containing always two vertices (\cite{95Feynman}, p.48).
The free field energy density will be positive in both cases.

For example the 4-scalar field $\varphi$ generated by the source
$T = \eta_{\mu\nu} T^{\mu\nu}$
(4-trace of the matter energy-momentum tensor)
can present both universal attraction and universal repulsion.
The corresponding massless field equation is:
\be \label{scal-att-rep}
-\varphi^{\,\,\,\;\;\;\,\,\,,\,\mu}_{(att)\,\,\,\mu} = + \lambda^2 T
\quad
\mbox{\rm and}
\quad
+\varphi^{\;\;\;\,\,\,\,\,\,\,,\,\mu}_{(rep)\,\,\,\mu} = +\lambda^2 T \,\,.
\ee
so the force acting on a slow moving test particle with mass $m_0$ will be
attractive or repulsive:  $\vec{F} = \pm m_0 \nabla \varphi$,
while the energy density for both attractive and repulsive fields is positive
\be \label{energy-att-rep}
T^{\,00}_{(\varphi) att} = T^{\,00}_{(\varphi) rep} = \kappa [(\dot{\varphi})^2 +
(\nabla \varphi)^2]
> 0 \,\,.
\ee

However in the case of simultaneous generation of both  attractive and
repulsive fields by a \emph{common source} $T^{\mu\nu}$ we get
the \emph{composite Lagrangian} with total kinetic term  containing
positive and negative parts, which does not mean negative energy
of the repulsion field.
The repulsive force partially compensates the attraction force,
so the resulting "effective" composite
potential, force, work and energy will be less than in the single irreducible tensor field component.

Intriguingly, such situation exactly happens in the case
of symmetric second rank tensor field   $\psi_{\mu\nu}$
generated   by the matter energy-momentum tensor $T_{\mu\nu}$.
Here we have composite reducible tensor potential
(direct sum of irreducible tensor and scalar parts) which is generated by
the common source for attractive and repulsive potentials.

\subsection{Conserved energy-momentum tensor of the field and
	positive localizable field's energy density}

Important aspect of the FGT is that the GW detection by LIGO-Virgo
observatories is consistent with the field nature of the
gravitational interaction. The detected gravitational waves
has positive localizable energy density as it should be for
other boson fields.

Inded, the basic initial principle of the quantum field theory states that
the action $\mathcal{A}$ (eq.\ref{act-fgt})
is the Poincar\'{e}-invariant 4-scalar in Minkowski space.
The conservation laws of energy-momentum and angular momentum,
i.e. the translational and Lorentz invariance of the Poincar\'{e}
covariant theory, is a consequence of the Noether's theorem:
existence and conservation of the energy-momentum tensor for any
physical field is the direct consequence of the symmetry of the
flat global Minkowski space.
So the ``prior geometry" of the
Minkowski space in the field theories has the advantage of
guarantee existence of the field energy-momentum, its localization and
its conservation.

The conserved canonical energy-momentum tensor (EMT) of the field  $\psi^{\mu\nu}$
is defined by the free field Lagrangian $\Lambda$:
\be \label{EMT-can}
T^{\alpha\beta}_{(can)}(x) =
\psi^{\,\mu\nu ,\,\alpha}(x)
\frac{\partial \Lambda}{\partial
	\psi^{\,\mu\nu}_{\,\,\,,\,\beta}(x)} - \eta^{\alpha\beta}\Lambda(x)
\quad
\mbox{\rm and}
\quad
T^{\alpha\beta}_{(can)\,\,,\,\alpha}(x) =0
\ee
with definite value at each point $x$.
The physically motivated EMT features are:
\begin{itemize}
	\item[$\bullet$]{\emph{$T^{\alpha\beta}(x)=T^{\beta\alpha}(x)$ -symmetry condition,
			necessary  for angular momentum conservation; }}
	\item[$\bullet$]{\emph{$T^{00}(x)  > 0$ - localizable field energy density,
			positive for both static and wave field, corresponding to the positive quant
			energy of integer spin fields  $E = h\nu$;}}
	\item[$\bullet$]{\emph{$T(x)= \eta_{\alpha\beta}T^{\alpha\beta}(x) =0$ - trace of the EMT is zero for massless particles;}}
	\item[$\bullet$]{\emph{the EMT from $\Lambda$ is defined not uniquely:
			$T^{\alpha\beta}=T^{\alpha\beta}_{(can)} + \partial_{\varrho}
			\Theta^{\varrho\alpha\beta}$ for
			$\Theta^{\varrho\alpha\beta} =- \Theta^{\varrho\beta\alpha}$. }}
\end{itemize}

Localization of the energy of the gravitational field means that the energy-momentum tensor $T^{\alpha\beta}_{(g)}(\vec{r}, t)$
is defined for any point $(\vec{r}, t)$ of the Minkowski spacetime
and its energy can be transformed (localized) in the
kinetic energy of test particles
(e.g. registration of gravitational wave amplitude
deeply inside its wavelength, as we actually have for LIGO events).

\subsection{Decomposition of the second rank symmetric tensor under
	the Lorentz group}
According to the group representation theory the symmetric second-rank tensor
$\psi^{\mu\nu}$ (10 components) is a reducible representation of the Lorentz group
which can be decomposed  into a direct sum of irreducible
representations: traceless 4-tensor $\textbf{T}$ (5 components) ,
4-vector $\textbf{V}$ (4 components)
and 4-scalar $\textbf{S}$ (1 component) fields (\cite{08Mag}, \cite{65Barnes}, \cite{65OgiPol}):
\be \label{10comp-rep}
\{\psi^{\mu\nu}\} = \{\textbf{T}\} \oplus \{\textbf{V}\}  \oplus
\{\textbf{S}\} \,.
\ee
Note the principal role of the Poincar\'{e} invariant scalar part
$S^{\mu\nu} = (1/4)\psi\eta^{\,\mu\nu}$
where $\psi = \eta_{\mu\nu}\psi^{\mu\nu}$ is the trace of the symmetric tensor
$\psi^{\mu\nu}$.

Below we shall consider special case when four supplement conditions
(gauge invariance and EMT conservation) delete
the 4-vector part of the eqs.(\ref{10comp-rep}), so the reducible
symmetric tensor will present the direct sum only two
irreducible parts -- the spin-2 and spin-0 fields.
In terms of spins this decomposition corresponds to following expression:
\be \label{2comp-spins}
\{\psi^{\mu\nu}\} = \{2\} \oplus [\{1\}  \oplus
\{0^{'}\}] \oplus  \{0\}\,
\quad
\Longrightarrow
\quad
\{\psi^{\mu\nu}\} = \{2\} \oplus  \{0\}
\ee
It means that our symmetric tensor field $\psi^{\mu\nu}$  can be
presented in the form:
\be \label{2+0-psi}
\psi^{\mu\nu} = \psi^{\mu\nu}_{\{2\}} +
\psi^{\mu\nu}_{\{0\}} = (\psi^{\mu\nu} - \frac{1}{4}\eta^{\mu\nu}\psi) +
\frac{1}{4}\eta^{\mu\nu}\psi =
\phi^{\mu\nu} + \theta^{\mu\nu}
\ee
and also the source  $T^{\mu\nu}$  (the matter EMT) will have the same
structure:
\be
\label{2+0-t}
T^{\mu\nu} = T^{\mu\nu}_{\{2\}} + T^{\mu\nu}_{\{0\}} =
(T^{\mu\nu} - \frac{1}{4}\eta^{\mu\nu}T)  +
\frac{1}{4}\eta^{\mu\nu}T\,.
\ee
Eqs.(\ref{2+0-psi}, \ref{2+0-t}) is the Poincar\'{e}-invariant decomposition of
both the initial symmetric tensor potential $\psi^{\mu\nu}$
and symmetric source $T^{\mu\nu}$ of this potential onto two irreducible parts.
This means that the true spin-2
field is only one part of this tensor  $-$ the 4-traceless field
$\psi^{\mu\nu}_{\{2\}} = \phi^{\mu\nu} = \psi^{\mu\nu} - \frac{1}{4}\eta^{\mu\nu}\psi$
(corresponds to spin-2 gravitons).
The second principally important part of the initial tensor potential
$\psi^{\mu\nu}$ is the 4-scalar spin-0 field
$\psi^{\mu\nu}_{\{0\}} = \theta^{\mu\nu} = (1/4)\psi \eta^{\mu\nu}$
(as we shall show below it
corresponds the repulsive component of the gravity phenomenon).

\section{ Lagrangians, propagators and amplitudes for massive gravitational field}

A massive 4-vector field has three degrees of freedom (dofs), namely two in the transverse modes and one in the longitudinal mode. Nevertheless, due to 4-current conservation, no external sources directly excite the helicity-0 mode of a massive spin-1 field, and the exchange amplitude between two conserved sources is the same in the limit $ m \rightarrow 0$ no matter whether the vector field is intrinsically massive and propagates three dofs or if it is massless and only propagates two modes. There is no difference between an exactly massive vector field and a massive one with arbitrarily small mass, which means that for the Proca mass term there is no mass discontinuity problem
(e.g. \cite{08Mag}, \cite{14Rham}).

Below we show that for the  symmetric tensor field $\psi^{\mu\nu}$
there is  analogous to ED smooth mass limit, if one takes into account
the decomposition of the reducible
field onto its intrinsic  irreducible spin-2 and spin-0 fields.
The helicity-0 mode now couple to the Poincar\`{e} invariant 4-trace of the
matter energy-momentum tensor and so generic sources will excite not only the 2 helicity-2 polarization of the \emph{graviton} but also a third helicity-0 polarization of the \emph{"repulson"} $\,$
(spin-0 Poincar\'{e} scalar field), which deliver a new possibility for
solution of the vDVZ mass discontinuity problem.

\subsection{Generalized Fierz-Pauli approach, gauge invariance and EMT conservation}

Let us consider Poincar\'{e}-invariant action (eq.\ref{act-fgt})
in the linear approximation.
We start with the Lagrangian for reducible symmetric second rank tensor field
$\psi^{\mu\nu}$ (which according to Wigner's theorem describes composition of three irreducible Poincar\'{e} invariant fields (eq.\ref{10comp-rep}))
with the generalized Fierz-Pauli mass term
(\cite{39FierzPauli}, \cite{95Feynman},  \cite{08RubTin}, \cite{08Mag}) :
\be \label{Lagr-tot-FP-mass}
\fl \Lambda_{(g)}=-{1\over{16\pi G}}\bigl[
2\psi_{\mu\sigma}{}^{,\,\mu}\psi^{\nu\sigma}{}_{,\,\nu}-
\psi_{\nu\sigma,\,\mu}\psi^{\nu\sigma,\,\mu}-
2\psi_{\nu\mu}{}^{,\,\nu}\psi^{,\,\mu}+
\psi_{,\,\nu}\psi^{,\,\nu} +
\frac{\hat{m}_g^2}{2}( a \psi_{\nu\sigma}\psi^{\nu\sigma} - b \psi^2)\bigr]
\,\,
\ee
where $\hat{m}_g=m_g c/\hbar$ is the inverse Compton length for the mass of the effective composite field,
and $a, b$ -  two arbitrary constants.
Taking into account the interaction Lagrangian (eq.\ref{ugi})
we get field equations (eq.\ref{field-eq}) in the form:
\be \label{FP-field-eq}
\fl - \psi^{\mu\nu,\,\sigma}{}_{\sigma} + \psi^{\mu\sigma,\,\nu}{}_{\sigma}{}
+ \psi^{\nu\sigma,\,\mu}{}_{\sigma} - \psi^{,\,\mu\nu}
- \eta^{\mu\nu} \psi^{\sigma\rho}{}_{,\,\sigma\rho}  + \eta^{\mu\nu} \psi^{,\,\sigma}{}_{\sigma}
- \frac{\hat{m}_g^2}{2}( a \, \psi^{\mu\nu} - b\, \eta^{\mu\nu}\psi)
= {8 \pi G \over c^2} T^{\mu\nu}\,
\ee
The trace of the tensor field equations (\ref{FP-field-eq}) gives:
\be
\label{feq-trace}
-2\psi^{,\,\sigma}_{\,\,\sigma} + 2\psi^{\sigma\varrho}_{\;\;\; ,\,\sigma\varrho}
+\frac{\hat{m}_g^2}{2} ( a  - 4 b)\psi
= -\frac{8 \pi G}{c^{2}} T\,.
\ee
In the massless limit the field equation (\ref{FP-field-eq}) is similar  to the linear approximation of Einstein's field equations and that is why there are many similarities between GRT and FGT in the weak field regime.

\paragraph{The role of the source energy-momentum conservation.}
Now let us take into account that according to the Noether's theorem the
source of the gravitational field   $T^{\mu\nu}$
is conserved (eq.\ref{EMT-can}), hence
we get from the field equation (eq.\ref{FP-field-eq}) the consistency condition:
\be \label{mass-consis-cond}
T_{\mu\nu}^{\,\,\,\,\,,\,\,\nu} =0
\quad
\Longrightarrow
\quad
\hat{m}_g^2(\psi_{\mu \nu }{}^{,\,\nu } - {b \over a}\, \psi_{,\,\mu }) =0
\ee
%

In the massless limit ($m_g \rightarrow 0$) the conservation law of the source in (eq.\ref{FP-field-eq}) also consistent with the gauge invariance of this field equations,
which are not changed by the following transformation of the potentials
(in the fixed reference frame):
\be
\label{gauge}
\psi^{\mu\nu} \Rightarrow \psi^{\mu\nu} + \lambda^{\mu,\,\nu}+
\lambda^{\nu,\,\mu}\,,
\quad
\mbox{\rm and}
\quad
\psi \Rightarrow \psi + 2\lambda^{\nu}_{\,\,\,,\,\nu}\,.
\ee
The four arbitrary functions $\lambda^{\mu}$ are consistent with the
four restrictions on the source EMT due to energy-momentum conservation.

\paragraph{The choice of the gauge conditions.}
Note that the source EMT conservation law $T^{\mu\nu}_{\,\,\,,\,\nu} =0$ is a
fundamental property of the Poincar\'{e} invariance of the action (eq.\ref{act-fgt}).
Also  decomposition of the symmetric tensor into three
irreducible parts (eq.\ref{10comp-rep}) is Poincar\'{e} invariant.
The EMT conservation law (\ref{mass-consis-cond})
together with the gauge invariance of the field equations
(\ref{gauge}) can be considered as
the spin restriction conditions for the composite structure of the initial
reducible symmetric second rank tensor $\psi^{\mu\nu}$, which
corresponds to deleting the source of the irreducible 4-vector field
(\cite{65OgiPol}, \cite{65Zakharov}).

The consistency condition eq.(\ref{mass-consis-cond}) demonstrates that
there is a possibility for having gauge conditions common for massive
and massless cases, which we will use below in the Hilbert-Lorentz form
(\cite{95Feynman}, \cite{61Thirring}, \cite{blago99}, \cite{08Mag})
\be \label{gauge-hl}
\psi^{\mu\nu}_{\;\;\; ,\,\,\nu} = \frac{1}{2}\, \psi^{\;,\,\,\mu}\,
\quad
\Rightarrow
\quad
\phi^{\,\mu\nu}_{\,\,\,,\,\,\nu} = \frac{1}{4} \, \psi^{\,,\,\mu}\,\,\,
\ee
so the constants $a =2$ and $b =1$
(\footnote{
	Fierz-Pauli gauge corresponds to the choice
	$a =b =1$, then  the 4-scalar $\psi$ is not
	a dynamical field, tensor and scalar parts are mixed,
 and there is no continuous transition
	from massive to massless cases.
}).
An important property of the Hilbert-Lorenz gauge (\ref{gauge-hl})
is that these conditions conserve decomposition on tensor and scalar parts.

With the gauge (\ref{gauge-hl}) the field equations (\ref{FP-field-eq})
are the Klein-Gordon equation :
\be \label{feq2-fgt}
\fl(\mathcal{D} - \hat{m}^2_g) \,(\psi^{\mu\nu} - \frac{1}{2}\eta^{\mu\nu}\psi)
=  \frac{8 \pi G}{c^{2}}   T^{\mu\nu} \,,
\quad
\mathrm{or}
\quad
(\mathcal{D} - \hat{m}^2_g) \,\psi^{\mu\nu}
=  \frac{8 \pi G}{c^{2}}   (T^{\mu\nu}
- \frac{1}{2}\eta^{\mu\nu}T)    \,,
\ee
and for the trace part $\psi = \eta^{\mu\nu}\psi_{\mu\nu}$
we get:
\be
\label{scalar-eq}
(\mathcal{D} - \hat{m}^2_g)\,
\psi (\vec{r},t)
= -  \frac{8 \pi G}{c^{2}} T (\vec{r},t)\,,
\ee
where d'Alambertian is
$\mathcal{D} = -\partial_{\mu}\partial^{\mu}= -(\cdot)^{\,\,,\,\mu}_{\,\,\,\,\,\mu}
= \left(\triangle - \frac{1}{c^2}\frac{\partial^2}{\partial t^2} \right)$.

\subsection{Lagrangians and field equations for composite symmetric tensor field }

Let us now compute the gravitational exchange amplitude between two sources
$T^{\,\prime}_{\mu\nu} \leftrightarrow T_{\mu\nu}$
in both the massive and massless fields cases.
For this we rewrite Lagrangian (eq.\ref{Lagr-tot-FP-mass}) and the field equations
(eq.\ref{FP-field-eq})  by using the Hilbert-Lorentz
gauge conditions (eq.\ref{gauge-hl})
and presenting of the field $\psi^{\mu\nu}$ and source $T^{\mu\nu}$
according to (eq.\ref{2+0-psi}, \ref{2+0-t}).
Then the total composite Lagrangian (eq.\ref{Lagr-tot-FP-mass})
will be the sum
of spin-2 and spin-0  free field Lagrangians:
\be \label{lagr-HL-2+0} \fl
\Lambda_{(g)} = \Lambda_{(2g)} + \Lambda_{(0r)} =
 \frac{1}{16 \pi G}
\left[\,  \phi_{\mu\nu , \,\sigma} \phi^{\mu\nu , \,\sigma}
+ \hat{m}^2_g \phi_{\mu\nu} \phi^{\mu\nu}
\right] \,
 - \frac{1}{64 \pi G} \left[ \psi_{,\,\nu} \psi^{\,,\,\nu}
+  \hat{m}^2_g \psi^2
\right] \,.
\ee
The interaction Lagrangian will have the form
$\Lambda_{(int)} = \Lambda_{(2int)} + \Lambda_{(0int)}$:
\be \label{ugi2-0}
\fl\Lambda_{(\mathrm{int})} =  -\frac{1}{c^{2}}
\, (\,\psi_{\{2\}\mu\nu} + \psi_{\{0\}\mu\nu})
(T^{\mu\nu}_{\{2\}} +    T^{\mu\nu}_{\{0\}}) =
-\frac{1}{c^{2}}\, ( \phi_{\mu\nu} T^{\mu\nu}_{\{2\}} +
\theta_{\mu\nu}T^{\mu\nu}_{\{0\}}\,).
\ee
Corresponding field equation for the traceless spin-2
field $\phi^{\mu\nu}$ is:
\be \label{feq-2-fgt}
\fl (\mathcal{D} - \hat{m}^2_g) \,\psi^{\,\mu\nu}_{\{2\}}
=  \frac{8 \pi G}{c^{2}}   T^{\mu\nu}_{\{2\}}
\qquad
\mathrm{\Rightarrow}
\qquad
(\mathcal{D} - \hat{m}^2_g) \,\phi^{\,\mu\nu}
=  \frac{8 \pi G}{c^{2}}  ( T^{\mu\nu} \, - \frac{1}{4}\eta^{\mu\nu} T )
\ee
The field equation for the scalar
part $\psi^{\,\mu\nu}_{\{0\}} = \theta^{\mu\nu} =\frac{1}{4}\eta^{\,\mu\nu}\psi$
will be:
\be \label{feq-0-fgt}
 (\mathcal{D} - \hat{m}^2_g) \,\psi^{\,\mu\nu}_{\{0\}}
= - \frac{8 \pi G}{c^{2}}   T^{\mu\nu}_{\{0\}} \,
\quad
\mathrm{\Rightarrow}
\quad
(\mathcal{D} - \hat{m}^2_g) \,\psi
= - \frac{8 \pi G}{c^{2}}  \, T
\ee
where $\mathcal{D}$ is the d'Alembertian.
The sum of the field equations (\ref{feq-2-fgt}, \ref{feq-0-fgt})
gives the composite field equation (\ref{feq2-fgt}).

In the massless limit
the field equations (\ref{feq2-fgt}) were used by Feynman
for the graviton's propagator calculation
(\cite{95Feynman}, eq.3.7.9). In particular  for the Feynman's
"bar operator" there are two relations:
\be \label{Tmn-2+0}
\overline{T^{\mu\nu}} = T^{\mu\nu} - \frac{1}{2}\eta^{\mu\nu}T =
(T^{\mu\nu} - \frac{1}{4}\eta^{\mu\nu}T) -
\frac{1}{4}\eta^{\mu\nu}T =
T^{\mu\nu}_{\{2\}} - T^{\mu\nu}_{\{0\}}
\ee
and
\be \label{Tmn-2+0} \fl
\,\,\,\,\,\,\,\,\,\,\,\,
\overline{\psi^{\mu\nu}} = \psi^{\mu\nu} - \frac{1}{2}\eta^{\mu\nu}\psi =
(\psi^{\mu\nu} - \frac{1}{4}\eta^{\mu\nu}\psi) -
\frac{1}{4}\eta^{\mu\nu}\psi =
\psi^{\mu\nu}_{\{2\}} - \psi^{\mu\nu}_{\{0\}} =
\phi^{\mu\nu} - \theta^{\mu\nu}
\,.
\ee
Hence our decomposition (\ref{2comp-spins})
onto irreducible traceless tensor and trace scalar
components gives physical explanation of the Feynman's bar operator
appearance due to the composite gravity source in equation
(\ref{feq2-fgt}) as the combination of the
traceless attractive (gravitons) and trace repulsive (repulsons)
forces.

In the Lagrangians
(eqs.\ref{Lagr-tot-FP-mass}, \ref{lagr-HL-2+0})
and the field equations (eqs.\ref{feq-2-fgt}, \ref{feq-0-fgt})
the tensor and scalar parts have opposite signs,
which does not mean negative energy of the scalar field
but reflects the opposite signs of the traceless tensor and
intrinsic scalar forces.

\subsection{Propagators for gravitons and repulsons}

According to Feynman's quantum field approach, the gravitational force
is a force between two energy-momentum tensors considered  matter
\cite{95Feynman},
\cite{blago99},
\cite{03Damour},
\cite{08Mag},
\cite{12Hinter},
\cite{14Rham}.

In the weak field limit the interaction energy is determined by
the single graviton exchange amplitude between two sources,
$T^{\mu\nu}_1 (x)$ and $T^{\mu\nu}_2 (y)$, which
takes the form
\be \label{Amplit-gen}
\mathcal{M} \sim \int d^4 x \int d^4 y
T^{\mu\nu}_1 (x)
G_{\mu\nu\alpha\beta}(x, y)
T^{\mu\nu}_2 (y)
\ee
where $G_{\mu\nu\alpha\beta}(x, y)$ is the corresponding propagator.

Using Feynman's path integral quantization method
for massless symmetric tensor field $\psi^{\mu\nu}$
(without decomposition),
the amplitude of the spin-2 particle exchange
 was derived by
(\cite{95Feynman}, \cite{65Zakharov},
\cite{blago99}, \cite{08Mag})
in the form
\be \label{amp-tot-massless} \fl \qquad
\mathcal{M}(k) =
\lambda^2 T'^{\rho\sigma}G_{\mu\nu\rho\sigma}
T^{\mu\nu} =
\lambda^2 T'_{\mu\nu}
(\frac{1}{k^2}) \overline{T^{\mu\nu}}
= \lambda^2 T'_{\mu\nu}
(\frac{1}{k^2})
(T^{\mu\nu} - \frac{1}{2}\eta^{\mu\nu}T )
\ee
with graviton propagator
\be \label{prop-psi-massless}
G_{\mu\nu\rho\sigma} (k) =
{1 \over 2} [\eta_{\mu\rho}\eta_{\nu\sigma} + \eta_{\mu\sigma}
\eta_{\nu\rho} -
 \eta_{\mu\nu}\eta_{\rho\sigma}]
({ 1 \over k^2 })\,.
\ee
However the reducible massless symmetric tensor field $\psi^{\mu\nu}$
cannot be extended smoothly to the massive case, which is
the vDVZ problem
(\cite{70vDV}, \cite{70Zakh},
\cite{blago99}, \cite{08Mag}).

In the irreducible decomposition picture,  where the potentials and the
source is presented by the irreducible traceless tensor plus scalar
components, there is natural possibility for transition to smooth
massive case  for both partial fields.
Indeed,  for massive traceless symmetric tensor field
$\psi^{\mu\nu}_{\{2\}} = \phi^{\mu\nu} $ and scalar field
$\psi^{\mu\nu}_{\{0\}} = \theta^{\mu\nu}$ there are simple
Lagrangians (eq. \ref{lagr-HL-2+0})
and equations of motion
(eqs.\ref{feq-2-fgt}, \ref{feq-0-fgt}) which have smooth limit
to the massless case.

Taking into account conservation
of energy-momentum tensor for the total composite field
together with additional condition of tracelessness for spin-2 parts
and the Hilbert-Lorentz gauge conditions (\ref{gauge-hl}) on the potentials
\be \label{emt-composit} \fl
\qquad
T^{\mu\nu}_{\,\,\,\,,\,\nu}=
(T^{\mu\nu}_{\{2\}} + T^{\mu\nu}_{\{0\}})_{\,,\,\nu} = 0\,,
\qquad
\mathrm{with}
\quad
\eta_{\mu\nu}T^{\mu\nu}_{\{2\}} =0\,,
\quad
\mathrm{and}
\quad
\phi^{\,\mu\nu}_{\,\,\,,\,\,\nu} = \frac{1}{4} \, \psi^{\,,\,\mu}\,,
\ee
and using the Lagrangian eq.(\ref{lagr-HL-2+0}) and field
equations (\ref{feq-2-fgt}, \ref{feq-0-fgt}) we get
amplitudes for simultaneous graviton and repulson exchange between
vertexes having conserved EMTs  ($\,T'^{\rho\sigma}, T^{\mu\nu}$)
(up to terms which vanish when contracted
with a conserved tensor):
\be \label{amp-tot-mass-2+0}
\fl
\mathcal{M}(k, m_g) =
\lambda^2 T'^{\rho\sigma}
G_{\mu\nu\rho\sigma}^{\,m}(k, m_g) T^{\mu\nu}
=\lambda^2
(T'_{\{2\}\,\mu\nu} + T'_{\{0\}\,\mu\nu})
(\frac{1}{k^2-m_g^2})
(T^{\mu\nu}_{\{2\}} - T^{\mu\nu}_{\{0\}})
\ee
where the total propagator is the sum of
the traceless massive graviton and massive scalar repulson
propagators:
\be \label{prop-2+0-mass}
\fl
G_{\mu\nu\rho\sigma}^{\,m} =
G^{\{2\}m}_{\mu\nu\rho\sigma} (k,m_g)
+ G^{\{0\} m}_{\mu\nu\rho\sigma} (k,m_g) =
{1 \over 2} \Biggl(
[\eta_{\mu\rho}\eta_{\nu\sigma} +
\eta_{\mu\sigma}\eta_{\nu\rho}] +
[-\eta_{\mu\nu}\eta_{\rho\sigma}] \Biggr)
\Biggl({ 1 \over k^2 -  m_g^2 } \Biggr)
\ee
The amplitudes for each field contain two parts $-$
instantaneous ($\propto 1/\kappa^2 $) and retarded
($\propto 1/(\omega^2 - \kappa^2 $), which describe static
and radiated fields.
In the massless limit $m_g \rightarrow 0$ we get smooth
transition to massless fields,
hence there is no mass discontinuity problem in FGT.

We should emphasize that gravitons and repulsons have united source
$T^{\,\mu\nu}$ through which they related according to Hilbert-Lorentz
gauge conditions and describe combination of the
attractive and repulsive forces. While outside the source these
fields are independent having helicity-2 and helicity-0 modes
with positive localizable energy density.

\subsection{Quantum nature of the Newtonian force and
gravitational wave experiments}

The quantum interpretation of the classical Newtonian gravity force
and testable relativistic gravity effects relate to the amplitudes
(eqs. \ref{amp-tot-mass-2+0})
calculation for corresponding EMT of gravity sources.
In the FGT linear weak field approximation the Newtonian gravity
is generated by a point particle EMT:
\be \label{EMT-p}
T^{\,\mu\nu}_{(p)} =  m_a c^2 \delta (\vec r -\vec r_a)
\sqrt{ 1-{v^2_a\over c^2}} u^\mu_a u^\nu_a
\ee
so the gravity source energy density and the trace are
\be \label{EMT-p-00-Tr}
\fl
T_{00} =  {m_a c^2
\over \sqrt{ 1-{v^2_a\over c^2}}}\,
\delta (\vec r -\vec r_a)
\quad
\mathrm{and}
\quad
T = m_a c^2 \delta (\vec r -\vec r_0)
\sqrt{ 1-{v^2_a\over c^2}}\,.
\ee
For a particle at rest the source EMT is the sum
of irreducible parts:
\be \label{EMT-p-2+0}
\fl
T^{\mu\nu}_{(p)} = T^{\mu\nu}_{(2)} + T^{\mu\nu}_{(0)} =
m c^2\,\, \delta (\vec r -\vec r_a)\,\,
[\,\,\frac{3}{4}\,\,
diag ( 1, { 1 \over 3} , { 1 \over 3} , { 1 \over 3} ) +
{ 1 \over 4}\,\, diag(1, -1, -1, -1) ]\,,
\ee
the "bar" source of gravitons and repulsons
according to field eqs.(\ref{feq-2-fgt}, \ref{feq-0-fgt}) is
\be \label{EMT-p-2+0}
\fl
\overline{T^{\mu\nu}}_{(p)} = T^{\mu\nu}_{(2)} - T^{\mu\nu}_{(0)} =
m c^2\,\, \delta (\vec r -\vec r_a)\,\,
[\,\,\frac{3}{4}\,\,
diag ( 1, { 1 \over 3} , { 1 \over 3} , { 1 \over 3} ) -
{ 1 \over 4}\,\, diag(1, -1, -1, -1) ]\,.
\ee

Hence the interaction energy due to exchange by graviton and repulson
between two particles $m_1$ and $m_2$
can be calculated by using propagator eq.(\ref{prop-2+0-mass}) and
corresponding normalization of the amplitude eq.(\ref{amp-tot-mass-2+0}).
For static massive traceless spin-2 field the interaction energy is:
\be \label{Newton-pot-2m}
\fl
U^{\{2\}}(r) = - { i 8 \pi G \over c^4} \int {d k^3 \over (2\pi)^3}
T^{00}_{\,1}
G^{\{2\}m}_{0000} (k,m_g)
T^{00}_{\,2}
 e^{ - i k r}
= - 2 \,
{ G m_1 m_2 \over r }\,\,
e^{-\frac{m_gc}{\bar{h}}\, r}
\ee
For static massive spin-0 repulson we get:
\be \label{Newton-pot-0m} \fl
U^{\{0\}}(r) =
- { i 8 \pi G \over c^4} \int {d k^3 \over (2\pi)^3}
T^{00}_{\,1}
 G^{\{0\}m}_{0000} (k,m_g)
T^{00}_{\,2}
 e^{ - i k r}
= +1\,
{ G m_1 m_2 \over r }\,\,
e^{ -\frac{m_gc}{\bar{h}}\, r}
\ee
The total interaction energy (attraction plus repulsion) will be given
by the sum $U^{\{2\}}+U^{\{0\}}$, which gives for the massless limit
($m_g \rightarrow$ 0) the Newtonian gravitational potential and total
tensor potential in Birkhoff's form:
\be \label{Birkhoff-pot}
\fl
U_N (r) = -\,\frac{Gm_1 m_2}{r}
\quad
\mathrm{and}
\quad
\psi^{\mu\nu} = \varphi_N \,\, diag(1,1,1,1)\,,
\quad
\mathrm{where}
\quad
\varphi_N = -\, \frac{Gm}{r} \, .
\ee
Equations of motion for particles in Birkhoff's  gravitational potential
and calculations of classical relativistic gravity effects were considered
in \cite{17Bar}, \cite{bar-oschep18}.

For a photon fly by the Sun
(energy density $T^{00}_{(em)} = \varepsilon\,\, \rightarrow \,\,
m_1 = \varepsilon/c^2$)
interacting with the Sun ($T_{(p)}^{00} = m_2c^2$) we have
exchange only  by spin-2 gravitons (trace of the electromagnetic
EMT equals zero). So the interaction energy
for the massless limit
($m_g \rightarrow$ 0)
will be:
\be \label{photon-pot}
\fl
U^{\{2\}}(r)
= - 2\,\, { G m_1 m_2 \over r }
\quad
\mathrm{and\,\,bending\,\,angle}
\quad
 \theta_{\mathrm{FGT}} = \theta_{\mathrm{GRT}} =
 2\theta_{\mathrm{N}} = \frac{4GM}{c^2 b}\,.
\ee
which corresponds to the observed bending angle by  the Sun.

Retarded part of the propagators eq.(\ref{prop-2+0-mass})
corresponds to gravitational wave
radiation for both 4-traceless tensor and 4-trace scalar irreducible parts
of the source EMT. The amplitude of the transversal helicity-2 GW is
the same in GRT and FGT (\cite{baryshev95},  \cite{17Bar}).
Both tensor (helicity-2)
and scalar (helicity-0) waves have localizable positive energy density.
The dynamical longitudinal mode (helicity-0 repulson) obeys
the wave equation (\ref{feq-0-fgt}), which in the massless limit is:
\be
\label{scalar-eq-GW}
\left(\triangle - \frac{1}{c^2}\frac{\partial^2}{\partial t^2} \right)
\psi (\vec{r},t)
= -  \frac{8 \pi G}{c^{2}} T (\vec{r},t)\,,
\ee
so it allows to perform an observational test
by measuring sky localization, signal amplitudes and polarizations of GW
using modern LIGO-Virgo facilities
(\cite{deRham17},
\cite{baryshev95}, \cite{bar-patu2001},
 \cite{17Bar},
\cite{fesik17}).

\section{ Conclusion}

The problem of linear vDVZ mass discontinuity can be naturally solved
in the frame of Feynman's quantum field theory approach if one takes
into account the composite structure of the symmetric tensor field
$\psi^{\mu\nu}$ and its conserved source -- the matter energy-momentum tensor $T^{\mu\nu}$.
In the quantum interpretation the Newtonian gravity force is
a result of the exchange by two kinds of virtual particles --
spin-2 gravitons (universal attraction) and spin-0 repulsons
(universal repulsion). Experimental/observational testing
of the quantum field gravitation theory is achievable with
advanced LIGO-Virgo gravitational wave detectors.


\end{document}